\documentclass[proof]{WileyASNA-v1}

\articletype{Conference Proceedings}

\received{30 November 2019}
\revised{13 May 2020}
\accepted{xxx 2020}

\begin{document}

\title{Double-dipping to Refine Stellar Rotation Periods }

\author[1]{Joanne Tan}

\author[1]{Gibor Basri*}

\authormark{ Tan \& Basri}

\address[1]{\orgdiv{Department of Astronomy}, \orgname{University of California, Berkeley}, \orgaddress{\state{California}, \country{United States}}}

\corres{*Gibor Basri. Department of Astronomy, University of California, Berkeley, CA 94720, USA. \\ \email{gbbasri@berkeley.edu}}

\abstract{We present a refined analysis of 15038 Kepler main sequence light curves to determine the stellar rotation periods. The initial period estimates come from an autocorrelation function, as has been done before. We then measure the duration of every intensity dip in the light curve, expressed as fractions of the initial rotation period estimate. These dip duration distributions are subdivided into several regions whose relation to each other helps determine which harmonic of the initial rotation period is most physically plausible. We compare our final rotation periods to those from \cite{McQuillan2014} and find that the great majority agree, but about 10\% of their periods are doubtful (usually twice as long as is most plausible). We are still refining our method, and will later extend it to more stars to substantially increase the sample of reliable stellar rotation periods.}


\jnlcitation{\cname{%
\author{Tan J}, and 
\author{Basri G}} (\cyear{2020}), 
\ctitle{Double-dipping to Refine Stellar Rotation Periods}, \cjournal{Astron. Nachr.}, \cvol{2020;00:1--4}.}

\maketitle

\footnotetext{\textbf{Abbreviations:} ACP, rotation period from the autocorrelation-based inferential method (autocorrelation period); DD, dip duration; McQ14, \cite{McQuillan2014}'s sample; MS, main sequence; RRR, region-to-region ratio; SDR, single-double ratio}

\section{Introduction}
\label{sec1}
Accurate measurements of stellar rotation periods are crucial to better our understanding of stellar activities. Despite the seemingly simple goal, a robustly accurate measurement of stellar rotation period is not so easy. To aid the realization of this goal, there are many ground- and space-based missions that have already and will continue to greatly increase the number of known stellar rotation periods. Here we focus on the Kepler mission \citep{Borucki2010} because of its long duration, and high cadence and precision. \cite{Santos2019}, \cite{Nielsen2013}, and \cite{Reinhold2013} and others have used stellar light curves to derive stellar rotation periods, each with different methodologies. This paper focuses on the work of \cite{McQuillan2014}, hereafter McQ14, in which an autocorrelation-based method was used to determine the rotation periods of 34030 Kepler main-sequence (MS) stars. 

In this paper, we focus on another characteristic of the stellar light curves - the distribution of intensity dip durations (DD). We measure the duration of each dip compared to the rotation period of the star. An autocorrelation-based inferential method (similar but not identical to the method of McQ14) is used to obtain the initial estimate of the period (hereafter called the ACP). We then separate the DD distribution into several regions, each covering a separate range of fractional periods, and obtain the numerical percentage of dips in each region. From these percentages, we define several region-to-region ratios (RRR) and utilize them to refine the rotation periods and test their reliability. This dip distribution analysis has proven useful to increase the robustness of the measurement of stellar rotation periods.

In Section \ref{sec2}, we delineate the criteria we have used to select our analysis sample. In Section \ref{sec3}, we describe our method in more detail. In Section \ref{sec4}, we discuss our findings and comparison to McQ14's results, as well as our future direction.

\section{Sample Selection}
\label{sec2}
\begin{figure}[h!]
\centering
\includegraphics[width=0.45\textwidth, height=0.37\textwidth]{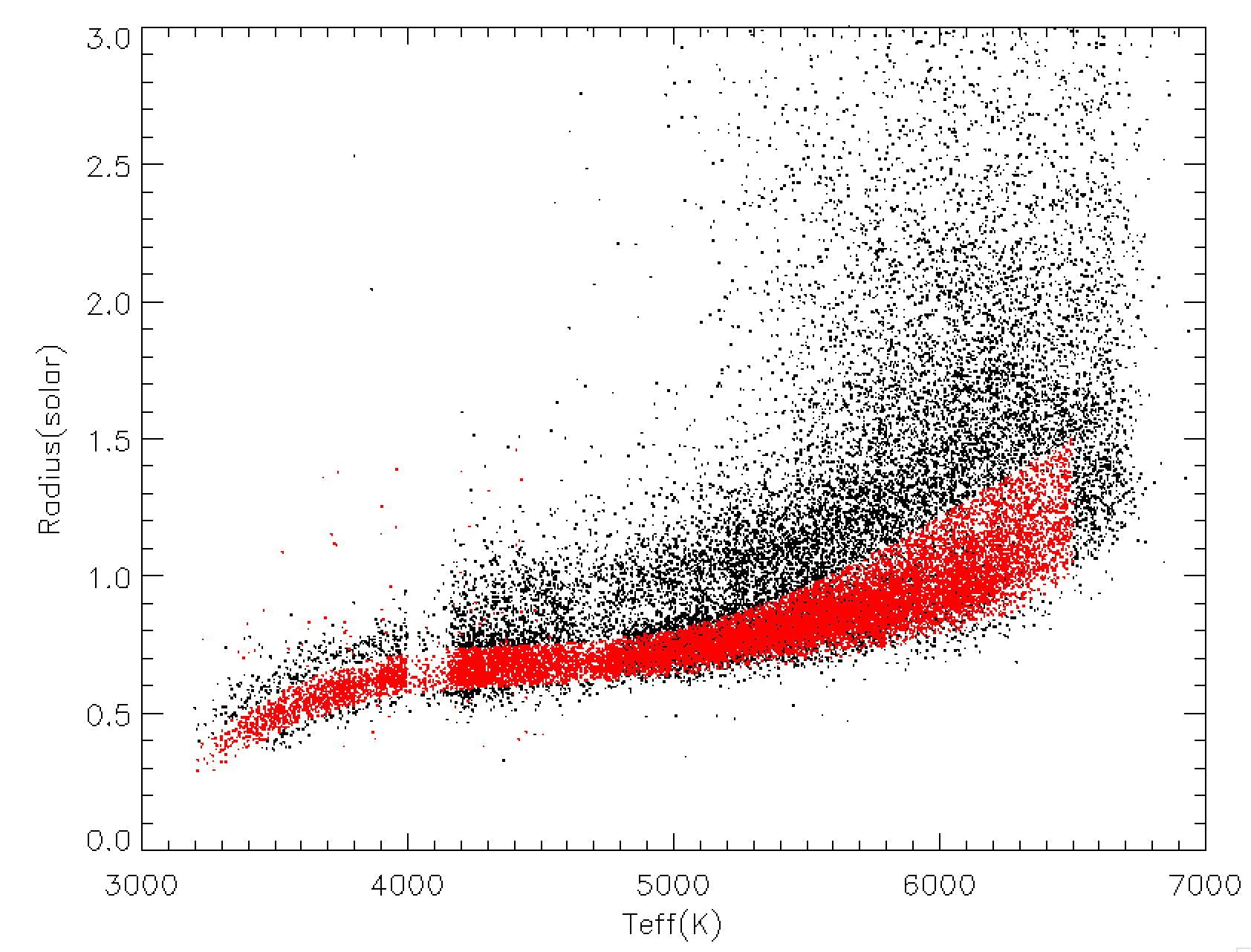}
\caption{\label{fig:1}
The full McQ14 sample of 34030 Kepler stars (black dots) showing the relation between stellar radius and stellar effective temperature, $T_{eff}$,  using Gaia DR2 stellar parameters. The red dots are the 15038 Kepler main sequence (MS) stars as defined in this paper.}
\end{figure}

McQ14 used the available stellar parameters from the Kepler Input Catalog, which have subsequently been found to have various inaccuracies. Our analysis uses the compilation of stellar temperatures and radii from \cite{Berger2018}, which in turn utilized the Gaia data release DR2, to select a sample of likely MS stars from the full McQ14 sample. We also took some care to eliminate confirmed and possible binaries. We then selected stars with at least eight quarters (out of the total of 17 quarters) of viable data, since not all stars were observed in all quarters. We omitted the data from Q1 and Q17 due to their short duration, and also rejected quarters which were different enough from the rest of the quarters that instrumental problems seem likely. We limited the brightness of the stars in our sample to $M_{Kep}$= 9 - 15.5, stellar temperature within the range $T_{eff}$ = 3000 - 6500K, and stellar radii within the expectation for single main sequence stars of each temperature. These criteria yield our full MS sample. From that we further restrict it to those that are also in the McQ14 sample. After all these filters, we obtained a sample of 15038 stars, which is roughly half that of McQ14, as shown in Fig. \ref{fig:1}. 

\section{Method}
\label{sec3}
Each Kepler long cadence light curve is first cleaned of bad quarters, and binned in time by a factor of 10 (because starspot variations cannot be nearly as short in duration as the half hour cadence of the Kepler data).

\subsection{Autocorrelation Analysis}
\label{sec3.1}
\begin{figure}[h!]
\centering
\includegraphics[width=0.45\textwidth, height=0.37\textwidth]{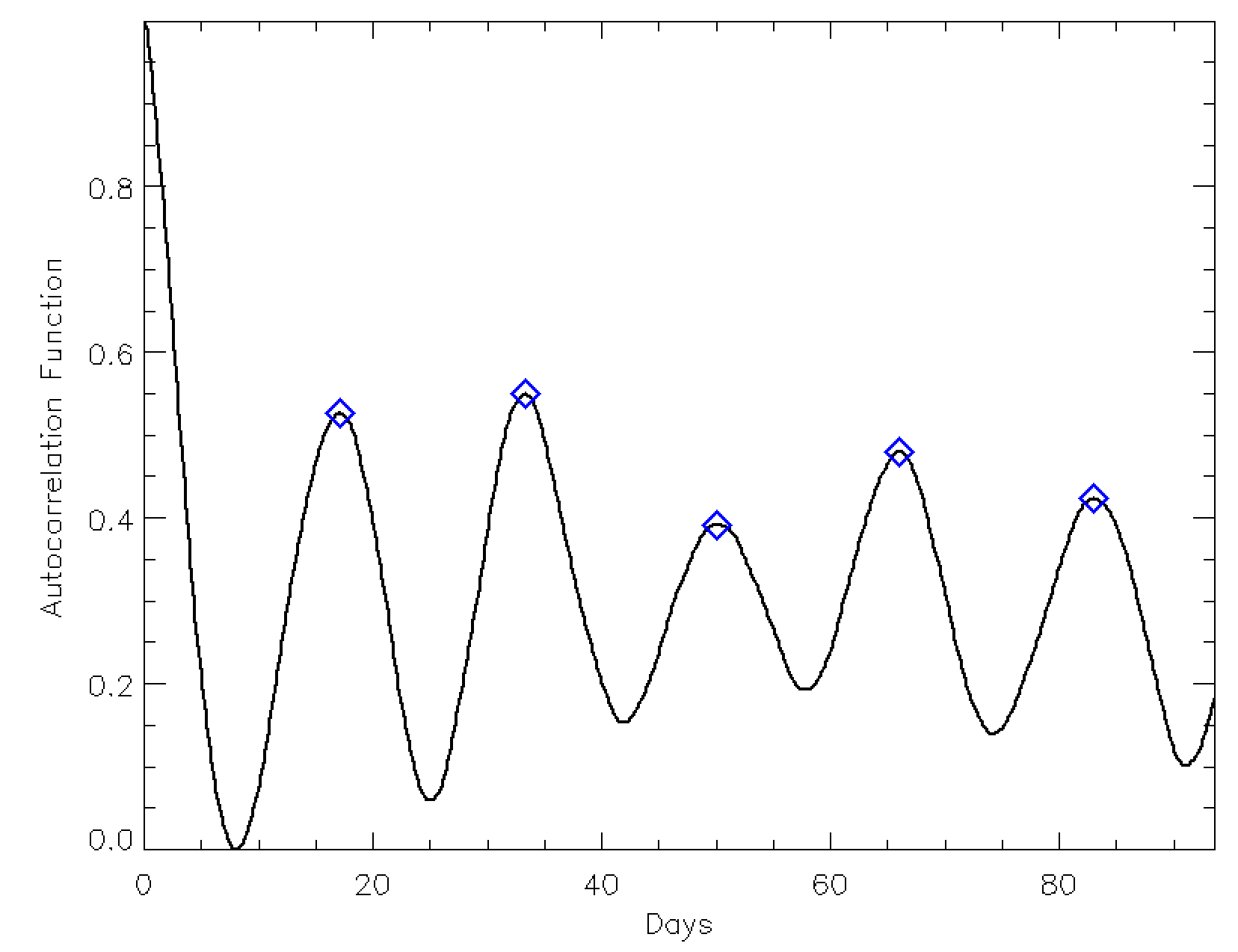}
\caption{\label{fig:2} The normalized autocorrelation function for the Kepler star KIC 6310092. The second and highest peak corresponds to both our initial period estimate (ACP) and the McQ14 period.
The first peak near 15 days corresponds to our final guess from the RRR selection criteria.}
\end{figure}

We perform an autocorrelation (AC) analysis on the stellar light curves in order to obtain AC peaks that provide information on the potential rotation period of the star. The AC function (often) exhibits repeating patterns of peaks with regular spacings. We focus on the first five peaks. In particular the highest of the first two peaks (refer to Fig. \ref{fig:2}) is chosen as the initial guess of the rotation period of the star (the "autocorrelation period" or ACP). Nevertheless, we then utilize other selection criteria as described in Section \ref{sec3.3} to reach our final choice.

The main issue with this method is distinguishing between periods and half-periods. It is very common for stars to switch between the two modes of either one dip per rotation or two (or more) dips per rotation. The extent and placement of single- and/or double-dip segments within stellar light curves is a confounding factor for most period determination routines. McQ14 utilized the fact that the autocorrelation function generated from the entire light curve may exhibit repeating patterns of alternating higher and lower peaks, which suggests the presence of a half-period signal. When this pattern is observed, they chose between the half- and full-periods by testing whether the second peak was higher than either the first or third, and then whether the first was at least 20\% lower than the second. The lower peaks were often taken to represent half-period signals. It is this procedure that we find leads to occasional errors in identifying the real period. However, like us, they also often utilize other selection criteria, which generally rectified this problem.

\subsection{Utilizing the Dip Duration (DD) Information}
\label{sec3.2}
\cite{Basri2018} studied in detail the behavior of single/double dipping modes in Kepler light curves. One must start with a guess of the period, which is supplied here by the ACP. The duration of every dip within a light curve is measured, then the duration is classified as single if it lasts 0.8 of the ACP or longer, or double if it lasts between 0.2-0.8 of the period. Any DD that lasts less than 0.2 of the period is considered as noise and neglected in the study. Using this form of classification, \cite{Basri2018} examined the mode durations and characterized them with the logarithmic ratio of the total time spent in the single mode to the total spent in the double mode. They termed this metric the single-double ratio (SDR), and found that it is a strong function of the stellar rotation period.

\begin{figure}[h!]
\centering
\includegraphics[width=0.48\textwidth, height=0.33\textwidth]{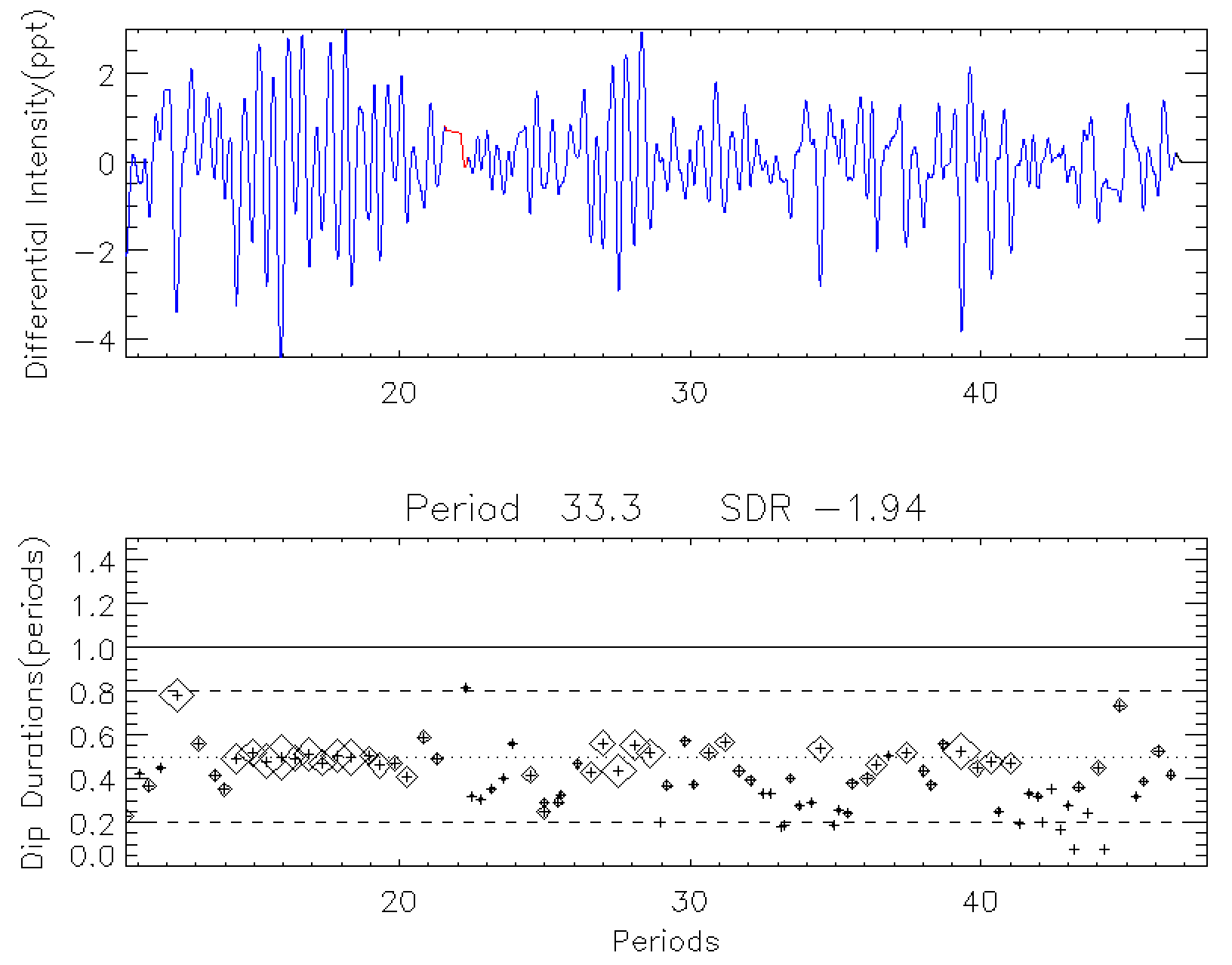}
\includegraphics[width=0.48\textwidth, height=0.33\textwidth]{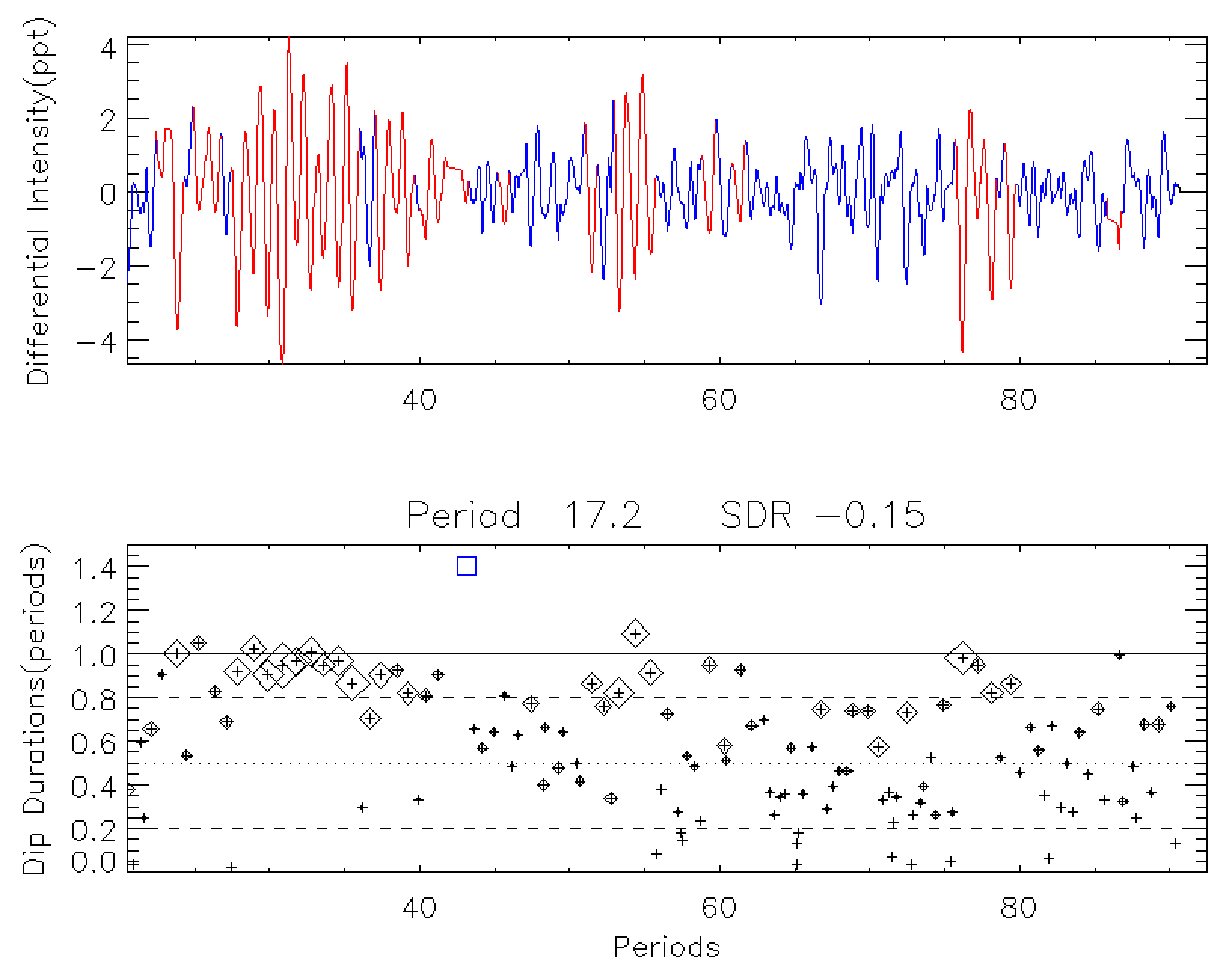}
\caption{\label{fig:3} The top panel shows the light curve for KIC 6310092 with the double-dipping mode (blue) and single-dipping mode (red) using the ACP/McQ14 period of 33.3 days. The panel below is the dip duration (DD) distribution for that case. The bottom two panels show the same things as above but with the period taken to be the RRR choice of 17.2 days. }
\end{figure}

\subsection{Region-to-Region Ratios (RRRs)}
\label{sec3.3}
In this study, we grouped the distribution of DD into several ranges of the fractional period (``regions"). Each region spans adjacent ranges of fractional periods: 0.2-0.5, 0.5-0.8, 0.8-1.2, and 1.2-2.0; we call them the R1, R2, R3, and R4 regions respectively. The number of dips in each region is then counted to obtain the percentage of dips in each region. We then compare various ratios of these regional percentages (RRR) to help us determine if our period determination method should pick the half-, double-, or initial period estimate. 

For a star in the double dipping mode, the dips should be somewhat evenly scattered above and below half a period, specifically in the R1 and R2 regions. In that case, the ratio of R1/R2 would be near unity. For a star in which R1 is much more populated than R2, it is more likely that half the ACP should be chosen as the final rotation period. That is because it is very difficult to maintain a starspot distribution that causes all the intensity dips to last under half a period (as our extensive spot modeling efforts have taught us). For a star dominated by the single dipping mode, the dips should be concentrated in the R3 region. That sort of distribution implies that the ACP should be kept as the final rotation period. 

One example that illustrates these points is shown in Fig. \ref{fig:3}, in which the star KIC 6310092 favors the shorter period. Our ACP was 33.3 days as the second AC peak is higher than the first, as shown in Fig. \ref{fig:2}. It is similar to the McQ14 period. When the ACP/McQ14 period of 33.3 days is used for the light curve and DD distribution, almost all dips are less than or just about half a period long, which is less physically plausible. Using the rotation period of 17.2 days as preferred by the RRR method, the DD distribution becomes much more physically reasonable, as the dips are scattered more evenly above and below half a period. Apparently, this star often switches between the single- and double-dipping mode. 

It is common for stars to have both single and double modes, which will populate R1, R2, and R3. Another reassuring characteristic of the correct period that we found from modeling and observations is that the median amplitude of single dips tends to be about twice as large as that of double dips. A rarer situation occurs when the dips are concentrated in the R4 and above regions - this means that the ACP is too short to reasonably characterize the stellar light curve (most intensity dips last longer than a period). In that case we choose double the ACP as the final rotation period.

After studying many cases in which the period is obvious, along with models in which we know the answer, we came up with several criteria on the relations between the RRRs that help us determine if the star's actual rotation period is half the ACP, double the ACP, or the original ACP. To determine an unknown period, we start with the ACP and those two harmonics. We then calculate the RRRs for all three cases. One caveat of the method is thus that we assume a star's rotation period is a multiple of the ACP. Given that, it is not surprising that we often agree with the values published by McQ14. From the additional selection criteria based on the RRRs, we generated the rotation periods for the 15038 Kepler stars in our sample. These are taken as the ``final" rotation periods for the rest of our analysis.

\section{Results}
\label{sec4}

\subsection{Comparison to McQuillan et al. (2014)} %\cite{McQuillan2014}}
\begin{figure}[h!]
\centering
\includegraphics[width=0.45\textwidth, height=0.37\textwidth]{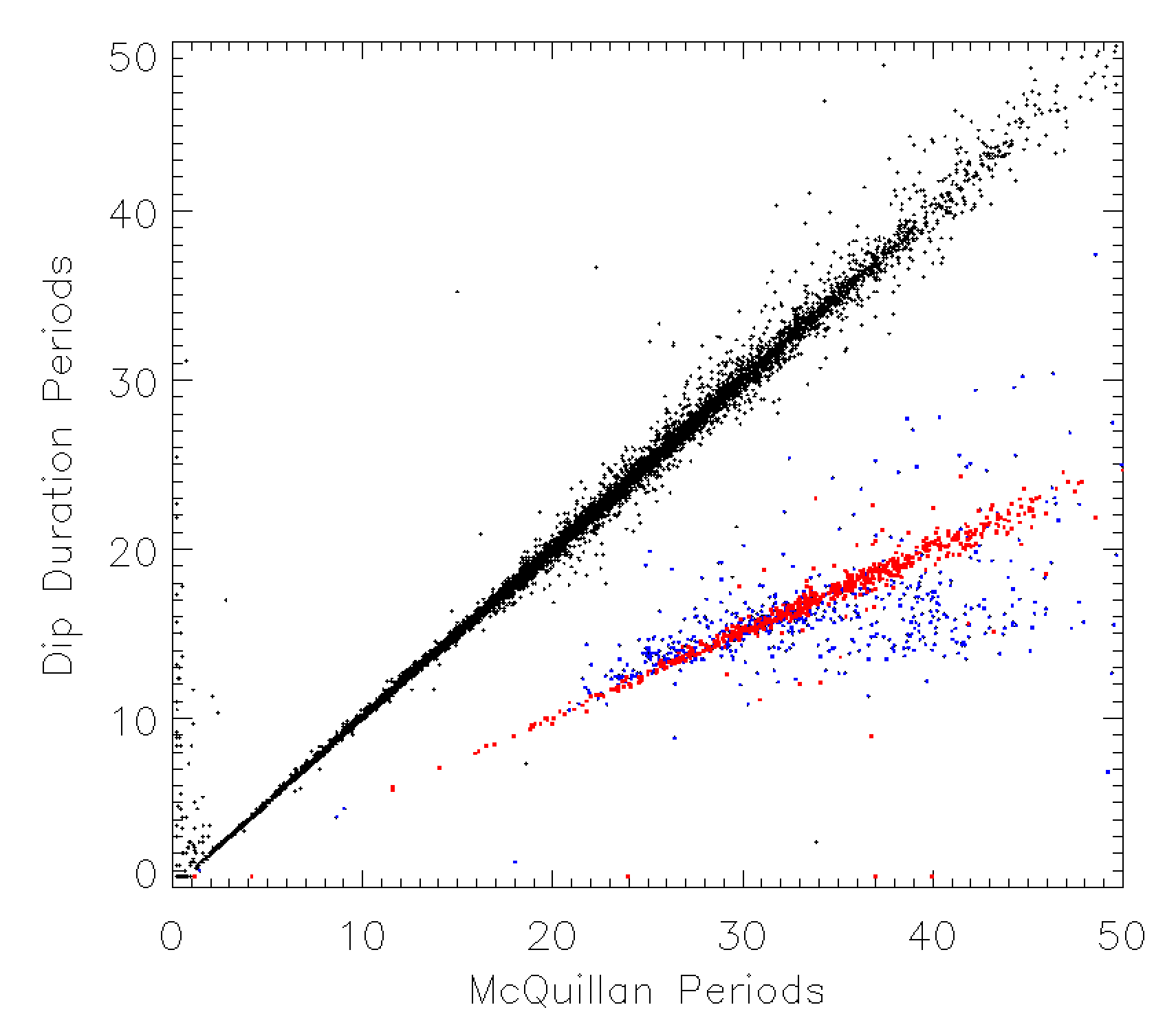}
\caption{\label{fig:4} Figure above shows the plot of our final rotation periods against the periods from McQ14. The points in black are the stars with rotation periods in close agreement with each other. The red points are stars with rotation periods from McQ14 about double those from our method. The blue points are cases where our period was not a harmonic of the period from McQ14.}
\end{figure}

\label{sec4.1}
We compare our final rotation periods to those published by \cite{McQuillan2014} in Fig. \ref{fig:4}. A large majority of the rotation periods from the two studies are very similar to each other as they lie close to the line of equality. Nevertheless, as seen in Fig. \ref{fig:4}, there is a population of stars in which the rotation periods from McQ14 are significantly larger than those in our study, often by a factor of close to 2. This discrepancy mainly stems from the different ways our methods interpret half- and full-periods. 

\begin{figure}[h!]
\centering
\includegraphics[width=0.48\textwidth, height=0.33\textwidth]{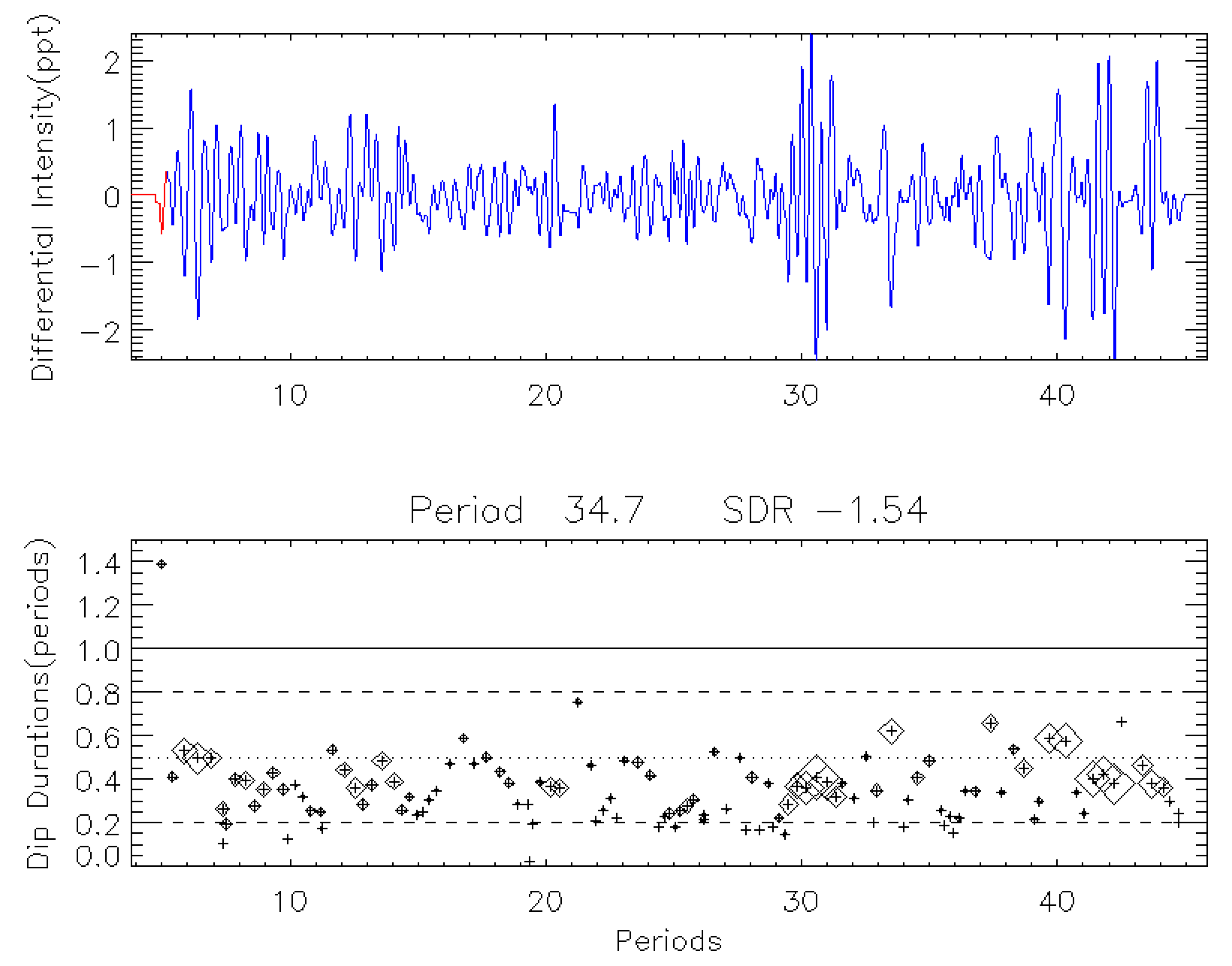}
\includegraphics[width=0.48\textwidth, height=0.33\textwidth]{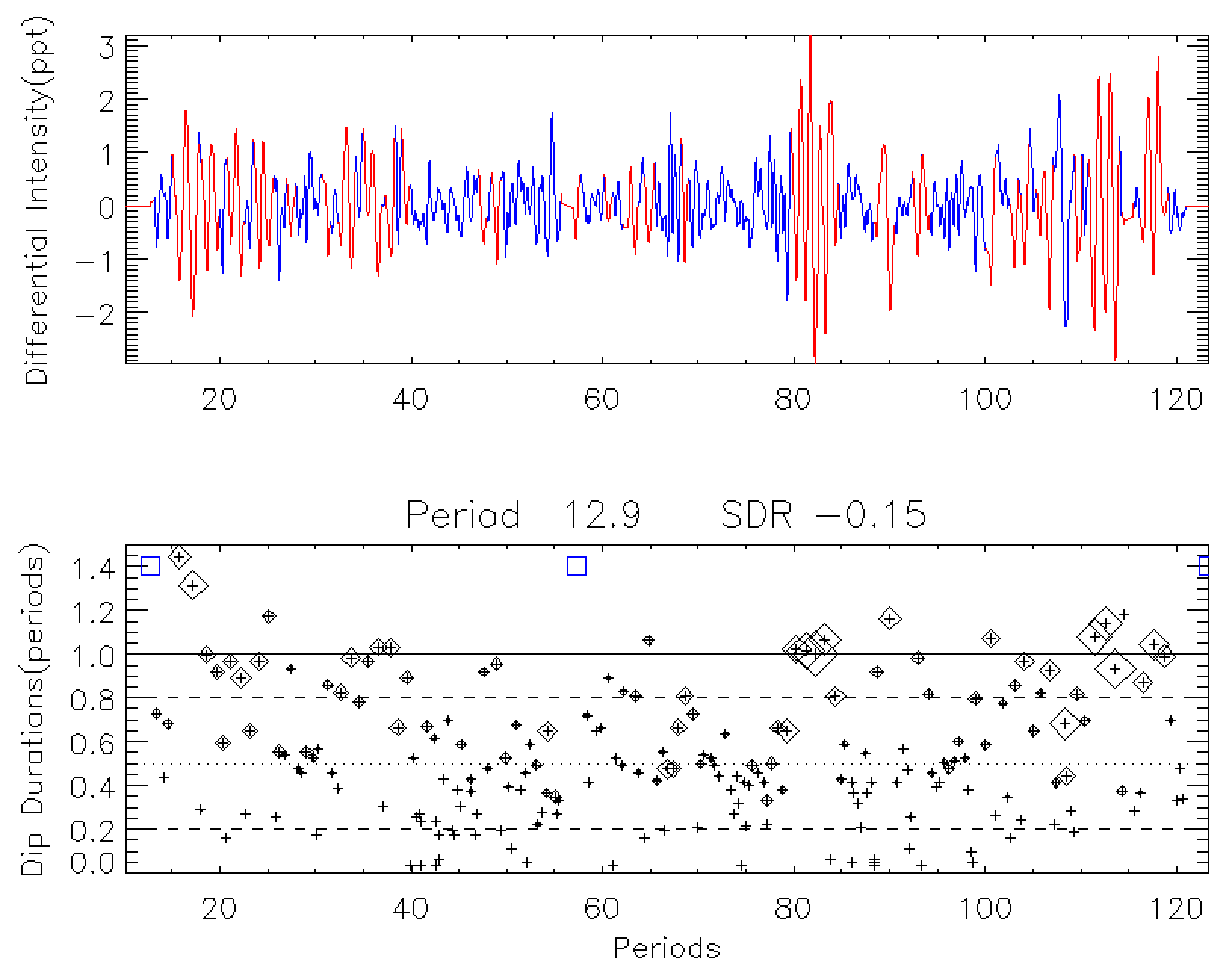}
\caption{\label{fig:6} Similar to the panels in Fig. \ref{fig:3}, the top two panels show the distribution of the light curve for KIC 9884284 in either double-(blue) or single-(red) dipping mode, and the DD distribution, when the rotation period used is 34.7 days (from McQ14). The bottom two panels are similar to the top two, but the rotation period used is instead 12.9 days, preferred by our ACP/RRR method.}
\end{figure}

Many of the disagreements (we claim) are interpretations by McQ14 of full-periods as half-periods. Using those McQ14 periods results in almost all of the DD being under half a period, while using our shorter final period leads to DD distributions that much better match the expectations garnered both from stars where the period is quite clear and from models. The reason for this is likely that stars switch between single and double modes with differing frequencies and lengths (as in Fig. \ref{fig:3} and Fig. \ref{fig:6}) and the distribution of each mode matters. Furthermore, the single mode peaks are typically twice the amplitude of the double mode peaks \citep{Basri2018}, giving them greater influence on the AC function.

\begin{figure}[h!]
\centering
\includegraphics[width=0.45\textwidth, height=0.37\textwidth]{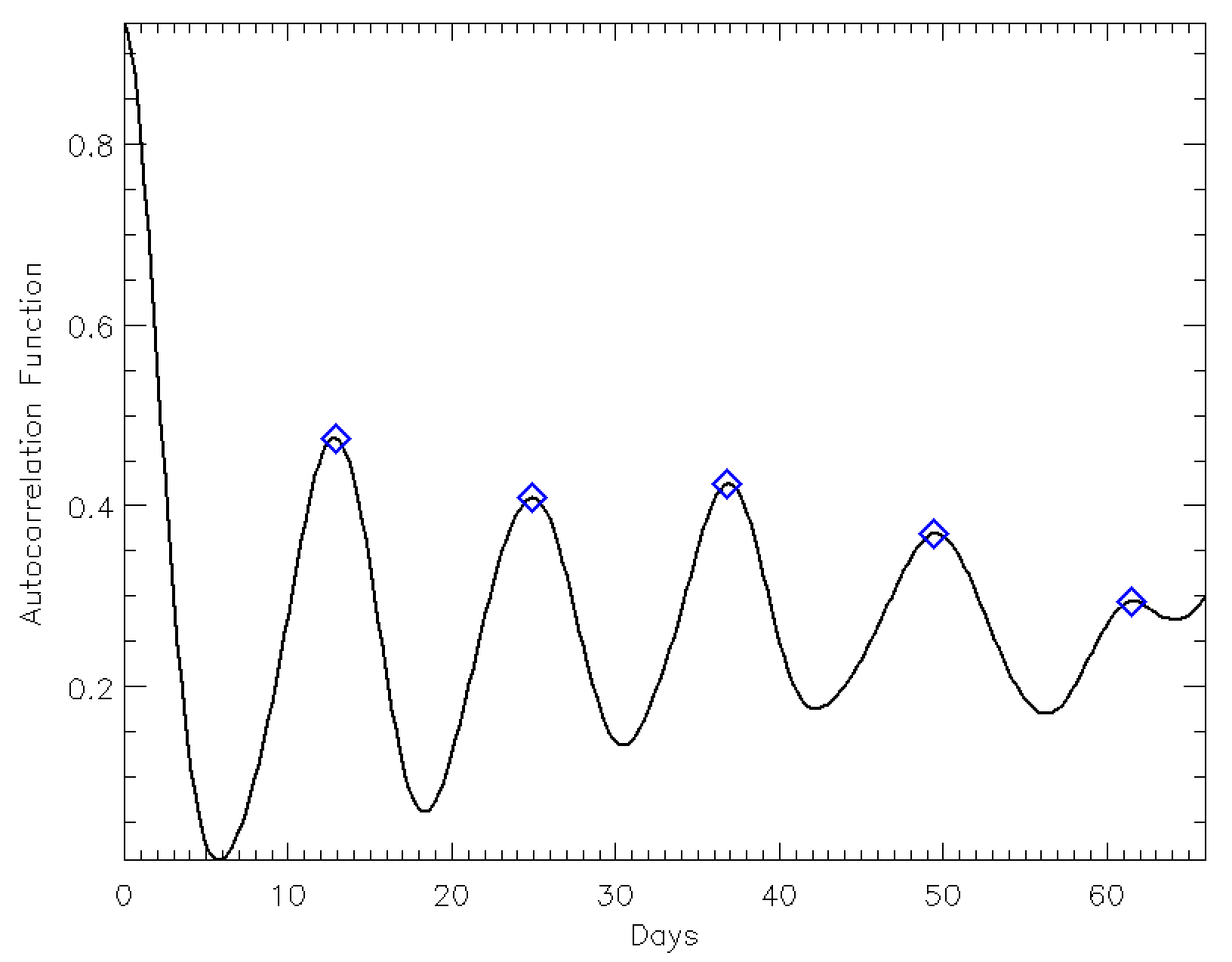}
\caption{\label{fig:5} The normalized autocorrelation function plot for the Kepler star KIC 9884284. The first and highest peak near 13 days corresponds to ACP/RRR period while the third peak corresponds to the McQ14 period.}
\end{figure}

Another example where the period choice differs is shown in Fig. \ref{fig:5} and Fig. \ref{fig:6}, which show the analysis for KIC 9884284. Fig. \ref{fig:5} shows the AC function and peaks. The period corresponding to the third peak was chosen by McQ14 as the period for this star instead of the first, presumably due to their other criteria. However, the DD plots shown in Fig. \ref{fig:6}, based on the two different rotation periods, indicate that the McQ14 period is likely to be incorrect since most dips in the light curve are less than half a period long when the McQ14 period is used. Using one third of that period, as selected by both ACP and RRR method, exhibits a more reasonable DD distribution and SDR, with dips scattered both above and below half a period. The amplitude of single dips are also larger than the double dips. These prompt us to believe that our solution is more reliable. Nearly half of the cases where McQ14 seems to choose double our period are characterized by the first two AC peaks having heights within 5\% of each other, whereas only less than 10\% of the full sample have such similar peak heights. In those cases, choosing the higher peak is a less reliable indicator of the true period because of the mode-switching between single/double can influence which peak ends up higher.

\subsection{Potential Improvements}
\label{sec4.3}
We are still in the process of investigating whether our new procedure has its own problems, and whether there are yet further diagnostics that would help with final reliability. There are definitely some problematic AC functions in which the spacing of the peaks is rather irregular and difficult to interpret. We are also looking into other characteristics of the stellar light curves and autocorrelation functions that may be useful in deciding the reliability of the ACP. This includes scrutinizing the regularity of the separations between dips in the autocorrelation functions, as well as the correlation between the range and SDR of the light curves and their correlations with rotation periods.

The problems worsen as the S/N of the light curve gets too low. Interpretations of noise as stellar signals can become fatal, especially for stars with small ranges of variability in their light curves. On the other hand, our methodology shows promise to reliably determine periods for less variable stars than were accepted by McQ14. We are exploring solutions that would help us identify and eliminate these issues. 

This study focused exclusively on Gaia MS stars that are in McQ14. In future work we will extend our analysis to two other groups of stars, namely Kepler Gaia MS stars that are not included in McQ14, and Kepler stars that are in McQ14 but are not Gaia MS stars. This will extend our sample from the current 15038 stars to roughly 60000 stars. This expansion will likely help in the quest to understand reliability, as well as greatly expand the list of known stellar rotation periods and extend it to stars with lower variability.

\section{Conclusions}
\label{sec5}
In this analysis of stellar light curves from 15038 Kepler MS stars, we find that constraints arising from the examination of intensity DD can be useful in refining the determination of rotation periods of stars. Global periodicity tests are sometimes insufficient for the task and it is better to utilize various methods to check for consistency in the rotation period determinations. Our refined method appears to increase the reliability of stellar rotation periods first determined from an autocorrelation analysis. We still find that a large majority of the periods from McQ14 are accurate, but about 10\% of them have come into question. This is a preliminary report, and we are continuing to refine our new methodology. As we expand our analysis to more stars, we expect to substantially increase the sample of reliable stellar rotation periods.

\section*{Acknowledgements}

This paper includes data collected by the Kepler mission. Funding for the Kepler mission was provided by the NASA Science Mission directorate. Most of the data presented in this paper were obtained from the Mikulski Archive for Space Telescopes (MAST). STScI is operated by the Association of Universities for Research in Astronomy, Inc., under NASA contract NAS5-26555.

\nocite{*}% Show all bib entries - both cited and uncited; comment this line to view only cited bib entries;
\bibliography{main}%

\begin{thebibliography}{}

\bibitem [\protect \citeauthoryear {%
Basri%
\ \BBA {} Nguyen%
}{%
Basri%
\ \BBA {} Nguyen%
}{%
{\protect \APACyear {2018}}%
}]{%
Basri2018}
\APACinsertmetastar {%
Basri2018}%
\begin{APACrefauthors}%
Basri, G.%
\BCBT {}\ \BBA {} Nguyen, H\BPBI T.%
\end{APACrefauthors}%
\unskip\
\newblock
\APACrefYearMonthDay{2018}{}{},
\newblock
\unskip
\newblock
\APACjournalVolNumPages{ApJ}{863}{}{190}.
\PrintBackRefs{\CurrentBib}

\bibitem [\protect \citeauthoryear {%
Berger%
, Huber%
, Gaidos%
\BCBL {}\ \BBA {} van Saders%
}{%
Berger%
\ \protect \BOthers {.}}{%
{\protect \APACyear {2018}}%
}]{%
Berger2018}
\APACinsertmetastar {%
Berger2018}%
\begin{APACrefauthors}%
Berger, T\BPBI A.%
, Huber, D.%
, Gaidos, E.%
\BCBL {}\ \BBA {} van Saders, J\BPBI L.%
\end{APACrefauthors}%
\unskip\
\newblock
\APACrefYearMonthDay{2018}{}{},
\newblock
\unskip
\newblock
\APACjournalVolNumPages{ApJ}{866}{}{99}.
\PrintBackRefs{\CurrentBib}

\bibitem [\protect \citeauthoryear {%
Borucki%
, Koch%
, Basri%
, Batalha%
\BCBL {}\ \BBA {} et al.%
}{%
Borucki%
\ \protect \BOthers {.}}{%
{\protect \APACyear {2010}}%
}]{%
Borucki2010}
\APACinsertmetastar {%
Borucki2010}%
\begin{APACrefauthors}%
Borucki, W\BPBI J.%
, Koch, D.%
, Basri, G.%
, Batalha, N.%
\BCBL {}\ \BBA {} et al.%
\end{APACrefauthors}%
\unskip\
\newblock
\APACrefYearMonthDay{2010}{}{},
\newblock
\unskip
\newblock
\APACjournalVolNumPages{Science}{327}{}{977}.
\PrintBackRefs{\CurrentBib}

\bibitem [\protect \citeauthoryear {%
McQuillan%
, Mazeh%
\BCBL {}\ \BBA {} Aigrain%
}{%
McQuillan%
\ \protect \BOthers {.}}{%
{\protect \APACyear {2014}}%
}]{%
McQuillan2014}
\APACinsertmetastar {%
McQuillan2014}%
\begin{APACrefauthors}%
McQuillan, A.%
, Mazeh, T.%
\BCBL {}\ \BBA {} Aigrain, S.%
\end{APACrefauthors}%
\unskip\
\newblock
\APACrefYearMonthDay{2014}{}{},
\newblock
\unskip
\newblock
\APACjournalVolNumPages{ApJS}{211}{}{24}.
\PrintBackRefs{\CurrentBib}

\bibitem [\protect \citeauthoryear {%
Nielsen%
, Gizon%
, Schunker%
\BCBL {}\ \BBA {} Karoff%
}{%
Nielsen%
\ \protect \BOthers {.}}{%
{\protect \APACyear {2013}}%
}]{%
Nielsen2013}
\APACinsertmetastar {%
Nielsen2013}%
\begin{APACrefauthors}%
Nielsen, M\BPBI B.%
, Gizon, L.%
, Schunker, H.%
\BCBL {}\ \BBA {} Karoff, C.%
\end{APACrefauthors}%
\unskip\
\newblock
\APACrefYearMonthDay{2013}{}{},
\newblock
\unskip
\newblock
\APACjournalVolNumPages{A\&A}{557}{}{L10}.
\PrintBackRefs{\CurrentBib}

\bibitem [\protect \citeauthoryear {%
Reinhold%
, Reiners%
\BCBL {}\ \BBA {} Basri%
}{%
Reinhold%
\ \protect \BOthers {.}}{%
{\protect \APACyear {2013}}%
}]{%
Reinhold2013}
\APACinsertmetastar {%
Reinhold2013}%
\begin{APACrefauthors}%
Reinhold, T.%
, Reiners, A.%
\BCBL {}\ \BBA {} Basri, G.%
\end{APACrefauthors}%
\unskip\
\newblock
\APACrefYearMonthDay{2013}{}{},
\newblock
\unskip
\newblock
\APACjournalVolNumPages{A\&A}{560}{}{A4}.
\PrintBackRefs{\CurrentBib}

\bibitem [\protect \citeauthoryear {%
Santos%
\ \protect \BOthers {.}}{%
Santos%
\ \protect \BOthers {.}}{%
{\protect \APACyear {2019}}%
}]{%
Santos2019}
\APACinsertmetastar {%
Santos2019}%
\begin{APACrefauthors}%
Santos, A\BPBI R\BPBI G.%
, García, R\BPBI A.%
, Mathur, S.%
\ et al.\end{APACrefauthors}%
\unskip\
\newblock
\APACrefYearMonthDay{2019}{}{},
\newblock
\unskip
\newblock
\APACjournalVolNumPages{ApJS}{244}{}{21}.
\PrintBackRefs{\CurrentBib}

\end{thebibliography}

\end{document}